\documentclass[aps,preprint,showpacs,preprintnumbers,amsmath,amssymb,prd]{revtex4}
\pdfoutput=1
\usepackage{amssymb}
\usepackage{graphicx}

\usepackage{graphics,psfrag}
\usepackage{graphicx,psfrag}
\usepackage{epic,eepic}
\usepackage{color}
\usepackage{epsfig}
\usepackage{latexsym}
\usepackage{pstricks}

\def\be{\begin{equation}}
\def\ee{\end{equation}}
\def\ba{\begin{eqnarray}}
\def\ea{\end{eqnarray}}

\begin{document}

\title{Light bending in radiation background}

\author{Jin Young Kim }
\email{jykim@kunsan.ac.kr}

\author{Taekoon Lee}
\email{tlee@kunsan.ac.kr}

\affiliation{Department of Physics, Kunsan National University,
Kunsan 573-701, Korea}


\begin{abstract}

We consider the velocity shift of light in presence of radiation
emitted by a black body. Within geometric optics formalism we
calculate the bending angle of a light ray when there is a gradient
in the energy density. We model the bending for two simplified
cases. The bending angle is proportional to the inverse square power
of the impact parameter ($\propto 1/b^2$) when the dilution of
energy density is spherically symmetric. The bending angle is
inversely proportional to the impact parameter ($\propto 1/b$) when
the energy density dilutes cylindrically. Assuming that a neutron
star is an isothermal black body, we estimate the order of magnitude
for such bending angle and compare it with the bending angle by
magnetic field.

\end{abstract}


\maketitle


Photons traveling in quantum electromagnetics (QED) vacuum in
presence of electrically neutral medium such as background
electromagnetic field, thermal environment, etc, have modified
lightcone condition, i.e., $ v \ne c$. The velocity shift can be
described as the index of refraction in geometric optics.

In classical optics a light ray can be bent if there is a gradient
in the refractive index. Previously we have computed the bending
angles of light when the refractive index is nonuniform by external
electric and magnetic fields\cite{kl1,kl2}. In this work we consider
the light bending when the vacuum is non-trivial by electromagnetic
radiation. Our purpose in this paper is to model the simplest case
of the light bending induced by thermal radiation of compact
astronomical objects.


Classically the interaction of a light ray with electric or magnetic
field is prohibited by the linearity of classical electrodynamics.
However in QED it is possible by the vacuum polarization that allows
the photon to exist as a virtual $e^+ e^-$-pair via which the
external field can couple. The first study on nonlinear effect
 in the presence of an external
electromagnetic field was performed by Euler and Heisenberg \cite{eulhei}.
The low energy effective lagrangian describing the physics of such nonlinear
interaction is described by \cite{eulhei,schwinger}, in SI units,
 \ba
 {\cal L} &=& -\frac{c^2 \epsilon_0}{4} F_{\mu\nu}F^{\mu\nu} +
 \frac{\alpha^2\hbar^3 \epsilon_0^2}{90m^4c}\left[(
 F_{\mu\nu}F^{\mu\nu})^2+\frac{7}{4} ( F_{\mu\nu}\tilde
 F^{\mu\nu})^2\right]   \nonumber   \\
  &=& \frac{\epsilon_0}{2} ({\bf E}^2 - c^2 {\bf B}^2)  +
 \frac{2\alpha^2\hbar^3 \epsilon_0^2}{45m^4c^5}\left[
 ({\bf E}^2 - c^2 {\bf B}^2)^2 + 7 c^2 ({\bf E} \cdot {\bf B} )^2
 \right].   \label{lagrangian}
 \ea

The first order correction to the speed of light in the presence of
external elctromagnetic field from the above lagrangian is given by
\cite{Bialynicka,adler,heyher,lorenci}
 \be
 \frac{v}{c} = 1-\frac{a\alpha^2\hbar^3 \epsilon_0}{45m^4c^3}
 \left[{\bf u\times E}/c + {\bf u\times (\bf u\times B)}\right]^2  ,  \label{velmag}
 \ee
where $\bf u$ denotes the unit vector in the direction of photon
propagation, and $a$ is a constant that depends on the photon
polarization. Dittrich and Gies \cite{ditgies,giesdit,gies99}
developed a general formula to find the velocity shifts and
refractive indices for soft photons traveling in QED vacuum modified
by external media based on an effective action approach. The light
cone condition under homogeneous media can be described by
 \be
 k^2 = Q < T^{\mu\nu} > k_\mu k_\nu ,   \label{lightcone}
 \ee
where $< T^{\mu\nu} >$ is the expectation of the energy-momentum
tensor in the modified vacuum and $Q$ is the so-called effective
action charge that depends on the parameters of the effective
action.

One can represent the light cone condition in terms
of velocity of light by choosing a certain reference frame and introducing
 \be
 {\bar k}^\mu = \left( \frac{k^0} {|{\bf k}|} , \hat{\bf k} \right) = ( v, \hat{\bf k} ) ,
    \label{phasevel}
 \ee
 where $v = k^0 / | {\bf k} |$ is the phase velocity.
 Then the equation (\ref{lightcone}) can be written as
 \be
 v^2 = 1 - Q < T^{\mu\nu} > {\bar k}_\mu {\bar k}_\nu .   \label{vsquare}
 \ee
When the correction is small, $Q <T^{00}> \ll 1$, the speed of light averaged over
 the propagation direction is given by
 \be
 v^2 = 1 -  \frac{4}{3} Q <T^{00}> = 1 -  \frac{4}{3} Q \rho , \label{vsquareaver}
 \ee
where $\rho$ is the energy density of general non-trivial QED vacuum.

The above formalism can be applied to all-loop effective actions.
In the weak field limit, the two-loop corrected velocity shift averaged over
polarization and propagation direction is obtained as \cite{ditgies}
 \be
 v = 1 - \frac{4 \alpha^2} {135 m^4}
 \left( 11 + \frac{1955}{36} \frac{\alpha}{\pi} \right)
 \left [ \frac{1}{2} ({\bf E}^2 + {\bf B}^2 ) \right ] .
 \ee
To the leading order (one-loop), this agrees with the result
calculated from Euler-Heisenberg lagrangian.

Let us calculate the velocity shift when a light ray passes the
non-trivial vacuum induced by the energy density of electromagnetic
radiation. The energy momentum tensor of incoherent radiation (real
photons) can be described by
 \be
 T^{\mu \nu} = \rho U^\mu U^\nu ,
 \ee
where $\rho$ is the energy density of photons emitted by radiation
background and $U^\mu$ is a null propagation vector in the direction
of radiation. If the radiation is spherically symmetric, $U^\mu=
(1,1,0,0)$ in spherical polar coordinate system. The light cone
condition (\ref{lightcone}) can be written as
 \be
 k^2 = Q \rho (U \cdot k)^2 = Q \rho (k_0 - k_r )^2 .
 \ee
The effective action charge $Q$ can be calculated from the box
diagram interaction \cite{Bialynicka}
 \be
 Q_{\pm} = - \frac{a_\pm \alpha^2 \hbar^3} {45 m^4 c^5},
 \ee
 where $a_+ = 14$ and $a_- = 8$ depending on the polarization.
 Taking the average over polarization, $Q_{\rm av} = - { 11 \alpha^2 \hbar^3}/{45 m^4 c^5}$,
 the velocity shift at one-loop level is given by
 \be
 \frac{v}{c} = 1 - \frac{ 11 \alpha^2\hbar^3 }{90 m^4c^5} \rho \left( 1 -
 \frac{k_r} { |{\vec k}|} \right )^2 .  \label{velrad}
 \ee


In geometric optics, one can calculate the bending of light ray if the index of refraction
is varying continually.
The trajectory of light ray for a continually varying index of
refraction can be written as \cite{kl1,kl2}
 \be
 \frac{d{\bf u}}{ds}=\frac{1}{n}({\bf u}\times {\nabla}n)\times {\bf u},
 \ee
where $s$ denotes the distance parameter
 of the light trajectory with $ds=|d\vec{\bf r}|$ and ${\bf u}={d\vec{\bf
 r}}/{ds}$ is the unit vector of the light ray.
 When the correction to the index of refraction due to the non-trivial vacuum is very small,
 the trajectory equation can be approximated to the leading order as
 \be
 \frac{d{\bf u}}{ds}=({\bf u}_0\times {\nabla}n)\times {\bf u}_0,
 \ee
 where ${\bf u}_0$ denotes the initial direction of the incoming ray.
For a ray coming in from $x=-\infty$ and moves to $+x$ direction,
 \be
 {\bf u}_0=(1,0,0),
 \ee
and defining ${\nabla}n \equiv (\eta_1,\eta_2,\eta_3)$, the
trajectory equations for $y(x)$ and $z(x)$ to the leading order are
given by
 \be
 \frac{d^2y}{dx^2}=\eta_2, ~~~~~~
 \frac{d^2z}{dx^2}=\eta_3.  \label{trajectory}
 \ee


Now we consider the trajectory of a light ray if there is a gradient
in the energy density of radiation. As a source of the lens, we
consider two simple cases when the gradient is spherically symmetric
or cylindrically symmetric. Let us consider the spherically
symmetric case first. Such case can be made by the dilution of
energy density thermally radiated from the surface of a compact
star. In general the temperature of any astronomical object may be
different for different surface points. For example, the temperature
of a magnetized neutron star on the magnetic pole is higher than
that on the equator due to magnetic field effects. However, for
simplicity, we will consider the mean effective surface temperature
as a function of radius assuming that a neutron is emitting energy
isotropically as a black body in steady state.

The energy density of free photons emitted by a black body at
temperature $T$ is given by the Stefan's law
 \be
 \rho = \frac{\pi^2}{15 \hbar^3 c^3} (k_B T )^4,
 \ee
 where $k_B$ is the Boltzmann's constant.
The dilution of energy density as a function of radius can be obtained from the conservation of
total radiation power, $L = 4 \pi r_0^2 \rho_0 =  4 \pi r^2 \rho$,
 \be
 \rho(r) = \rho_0 \frac{r_0^2 }{r^2} ,
 \ee
where $r_0$ is the radius of a spherical black body and $\rho_0$ the
energy density of a spherical black body with surface temperature
$T_0$ at $r_0$. The index of refraction, to the leading order, is
given by
 \be
 n(r) = \frac{c}{v(r)} = 1 + \frac{11 \pi^2 \alpha^2}{1350}
 \left ( \frac{ k_B T_0 } {mc^2} \right )^4 \frac{r_0^2}{r^2}
 \left( 1 - \frac{k_r} { |{\vec k}|} \right )^2 .
   \label{nindexsph}
 \ee
The electron mass energy $m c^2$ can be replaced by $k_B T_c$ with
the critical temperature of QED defined as $T_c = mc^2 / k_B = 5.94 \times 10^9 \rm K$.

Taking the direction of the incident ray as $+ x$ axis on the $xy$
plane ($ r= \sqrt{x^2 +y^2}$), the index of refraction can be
written as
 \be
 n(r) = 1 + \frac{11 \pi^2 \alpha^2}{1350}
 \left ( \frac{ T_0 } {T_c} \right )^4 \frac{r_0^2}{r^2}
 \left( 1 - \frac{x}{r} \right )^2 ,
   \label{nindexsphx}
 \ee
and the trajectory equation at the leading order is obtained as
 \be
 y^{\prime \prime} =
 \frac{22 \pi^2 \alpha^2}{1350}
 \left ( \frac{ T_0 } {T_c} \right )^4 r_0^2 y
 \left(  -\frac{3}{r^4} + \frac{3 x}{r^5} +\frac{2 y^2}{r^6} \right )
 . \label{trajecsph}
 \ee
For the incoming photon with the impact parameter $b$, the initial
condition reads
 \be
 y(-\infty) = b, ~~~~y^{\prime} (-\infty) = 0.
 \ee
 Integrating Eq. (\ref{trajecsph}) with $y=b$
 for the leading order solution, we obtain
 \be
 y^{\prime } (x) = - \frac{22 \pi^2 \alpha^2}{1350}
 \frac{ T_0^4 } {T_c^4} \frac{r_0^2}{ b^2}
 \left[ \frac{3}{4} \left( \tan^{-1} \frac{x}{b} + \frac{\pi}{2} \right)
         + \frac{b^3}{ (b^2 +x^2 )^{\frac{3}{2}} }
         + \frac{ 3 bx }{ 4 (b^2 +x^2 )}
         - \frac{ x b^3}{ 2 (b^2 +x^2 )^2}
                                  \right] ,    \label{yprimesph}
 \ee
 \be
 y (x) = b \left[ 1 -
  \frac{22 \pi^2 \alpha^2}{1350}
 \frac{ T_0^4 } {T_c^4} \frac{r_0^2}{b^2}
 \left\{ \frac{3}{4} \frac{x}{b}
  \left( \tan^{-1} \frac{x}{b} + \frac{\pi}{2} \right)
         + \frac{b^2}{ 4 (b^2 +x^2 ) }
         + \frac{x}{ \sqrt{b^2 +x^2 } } + \frac{7}{4} \right\} \right] .
  \label{ysph}
 \ee
The total bending angle $\varphi_{\rm sph}$ can be obtained from
$y^\prime (\infty) = \tan \varphi_{\rm sph} \simeq \varphi_{\rm
 sph}$,
 \be
 \varphi_{\rm sph} = -
 \frac{11 \pi^3 \alpha^2}{900}
 \frac{T_0^4 } {T_c^4} \frac{r_0^2} { b^2} .
 \label{varphisph}
 \ee


Now we consider the cylindrically symmetric case. Taking the axis of
cylinder as $z$-axis, from the conservation of the radiation power,
the dilution of energy density is given by
 \be
 \rho(r) = \rho_0 \frac{r_0}{r} ,
 \ee
where $r = \sqrt{x^2 +y^2}$ and $r_0$ is the radius of the
cylindrical black body. The refractive index can be written as
 \be
 n(r) = 1 + \frac{11 \pi^2 \alpha^2}{1350}
 \left ( \frac{ T_0 } {T_c} \right )^4 \frac{r_0}{r}
 \left( 1 - \frac{x}{r} \right )^2 ,
   \label{nindexcyl}
 \ee
 For a ray moving to the $+x$ direction in the $xy$ plane,
 the leading order trajectory equation can be written as
 \be
 y^{\prime \prime} =
 \frac{11 \pi^2 \alpha^2}{1350}
 \left ( \frac{ T_0 } {T_c} \right )^4 r_0 y
 \left(  -\frac{4}{r^3} + \frac{4 x}{r^4} +\frac{3 y^2}{r^5} \right )
 .      \label{trajeccyl}
 \ee
For the impact parameter $b$, the solution is obtained as
 \be
 y^{\prime } (x) = - \frac{11 \pi^2 \alpha^2}{1350}
 \frac{ T_0^4 } {T_c^4} \frac{r_0}{ b}
 \left[ \frac{4}{3} \left(  \frac{x}{ \sqrt{b^2 +x^2 } }
                + 1 \right)
     - \frac{1}{3}  \frac{b^2 x}{ (b^2 +x^2 )^{\frac{3}{2}} }
      + \frac{ 2 b^2}{ b^2 +x^2 }
                                  \right] ,    \label{yprimecyl}
 \ee
 \be
 y (x) = b \left[ 1 -
  \frac{11 \pi^2 \alpha^2}{1350}
 \frac{ T_0^4 } {T_c^4} \frac{r_0}{b}
 \left\{  2 \left( \tan^{-1} \frac{x}{b} + \frac{\pi}{2} \right) +
  \frac{4}{3} \frac{\sqrt{b^2 +x^2} +x}{b}
         + \frac{1}{3} \frac{b}{\sqrt{b^2 +x^2}}
          \right\} \right] .
  \label{ycyl}
 \ee
The total bending angle obtained from $y^\prime (\infty)$ is
  \be
  \varphi_{\rm cyl} = - \frac{44 \pi^2}{2025} \alpha^2 \frac{T_0^4 }{T_c^4}
  \frac{r_0}{b}.  \label{varphicyl}
  \ee


Note that the bending angles in (\ref{varphisph}) and
(\ref{varphicyl}) depend on the factor $(T_0 /T_c )^4$. Since the
maximal value of $T_0$ on ground laboratory and a normal star like
sun is of the order $T_0 \sim 10^3 \rm K$ while
 $T_c \sim 10^9 \rm K$,
detecting the bending by radiation on ground experiment or in the
neighborhood of a normal star seems very difficult. As an
application to possible real physical phenomena, let us consider the
light bending by a magnetized neutron star since the surface
temperature of neutron stars are pretty high and there is no
atmosphere to prevent the propagation of light ray.

For a magnetized neutron star the light bending can also occur by both the magnetic field and
gravitation. The bending by gravitational field is well-known from
general relativity
 \be
 \varphi_{\rm grav} = \frac{4G {\cal M}}{bc^2}.  \label{bendingG}
 \ee
The bending by magnetic field can be calculated with the index of
refraction obtained by Euler-Heisenberg lagrangian \cite{denisov,
denisov01}. The bending by a magnetic dipole generally depends on
the orientation of the magnetic dipole relative to the direction of
the incoming ray and we have computed a general formula on the
bending angles before \cite{kl2}. We consider the case when the ray
is passing the axis of dipole where the bending is maximal
 \be
 \varphi_{\rm mag} =
 \frac{ 41 \pi}{3 \cdot 2^7} \frac{a\alpha^2 \epsilon_0 c \hbar}{ e^2 }
 \frac{B_0^2}{B_c^2} \frac{r_0^6}{b^6} ,
 \label{bendingB}
 \ee
 where $a= 8$ or 14 depending on the polarization of the photon,
 $B_0$ is the magnetic field at the surface of neutron star, and
  $B_c$ is the critical magnetic field of QED defined by
 $B_c = m^2c^2/e \hbar = 4.4 \times 10^9 {\rm T}$.

Note that, from $ \varphi_{\rm mag} \propto 1 /b^6$ and $
\varphi_{\rm sph} \propto 1 /b^2$, the bending by magnetic field is
dominant at short distance while the bending  by the dilution of
real photon density is dominant at long distance. The power
dependence on the impact parameter, surface temperature, and
magnetic field is imprinted from the energy density through the
index of refraction $\nabla n \propto \nabla u$. For the bending by
a magnetic dipole, $u \propto B_0^2 /r^6$, and for the bending by a
spherical black body, $ u \propto T_0^4/r^2$.

To do an order-of-magnitude estimation, we consider the possible
bending angles as functions of impact parameter. The mass of neutron
star is of the order of solar mass so we take
 ${\cal M} \sim {\cal M}_{\rm sun} = 2 \times 10^{30} {\rm kg}$.
 We take the radius of neutron star as $r_0 = 10 {\rm
 km}$.
Most of the neutron stars possess surface magnetic field of the
order $B_0 = 10^4 - 10^9 \rm T$. Of course there are neutron stars
with the surface magnetic field above the QED critical limit known
as the magnetars. However, we do not consider such extremely strong
magnetic field since the calculation in (\ref{bendingB}) is based on
the Euler-Heisenberg lagrangian. Thus we consider the surface
magnetic field up to the order of $10^8 \rm T$ so that we take $B_0
/B_c \sim 10^{-1}$ as the upper bound. The surface temperature of
neutron stars is estimated as $ T_0 \le 10^6 \rm K$ \cite{page1,
page2, potekhin} and we take $ T_0 / T_c \sim 10^{-3}$ as the upper
bound of the temperature. The bending angles  for the above values
are estimated as
 \be
 \varphi_{\rm grav} \sim {\cal O} (10^{-1} ) \times \frac{r_0}{b} ; ~~
 \varphi_{\rm mag} \sim {\cal O} (10^{-5} ) \times
 \left( \frac{r_0}{b} \right)^6 ; ~~
 \varphi_{\rm rad} \sim{\cal O} ( 10^{-17} ) \left( \frac{r_0}{b} \right)^2 .
 \ee
 The bending by magnetic dipole field dominates the bending by radiation for $b < 10^3 r_0$,
 while the bending by radiation dominates the magnetic bending for $b > 10^3 r_0$.
 However, both bending angles are still small compared with
 the gravitational bending.

\acknowledgements{ We would like to thank T. Y. Koo, Y. Yi, and M. K. Park for
discussion and help. This research was supported by Basic Science
Research Program through the National Research Foundation of Korea
(NRF) funded by the Ministry of Education, Science and Technology
(12A12840581, 2012R1A1A2044543).}

\end{document}